# Sex differences in attitudes towards online privacy and anonymity among Israeli students with different technical backgrounds

**Maor Weinberger, Maayan Zhitomirsky-Geffet and Dan Bouhnik.**

***Introduction.*** *In this exploratory study, we proposed an experimental framework to investigate and model male/female differences in attitudes towards online privacy and anonymity among Israeli students. Our aim was to comparatively model men and women's online privacy attitudes, and to assess the online privacy gender gap.*
***Method.*** *Various factors related to the user's online privacy and anonymity were considered, such as awareness of anonymous threats made online, concern for protecting personal information on the Internet, online privacy self-efficacy, online privacy literacy and users' tendency to engage in privacy paradox behaviour, i.e., personal data disclosure despite the awareness of anonymity and privacy threats.*
***Analysis.*** *A user study was carried out among 169 Israeli academic students through a quantitative method using closed-ended questionnaires. The subjects' responses were analysed using standard statistical measures. We then proposed a summarized comparative model for the two sexes' online privacy behaviour.*
***Results.*** *We found that a digital gap still exists between men and women regarding technological knowledge and skills used to protect their identity and personal information on the Web. Interestingly, users' tendency to engage in privacy paradox behaviour was not higher among men despite their higher level of technological online privacy literacy compared to women.*
***Conclusions.*** *Women's relatively high online privacy self-efficacy level and their low awareness of technological threat do not match their relatively low technological online privacy literacy level. This leads to a lower ability to protect their identity and personal information as compared to men. We conclude that further steps should be taken to eliminate the inter-gender technological gap in online privacy and anonymity awareness and literacy.*

## Introduction

Anonymity is defined as '*the state of being not identifiable within a set of subjects*' (Pfitzmann and Köhntopp, 2001, p. 3). The desire for anonymity leads to the highest level of privacy, since it ensures the individual's ability to keep and protect his or her identity. Despite the fact that there are a variety of tools available to obscure user identity and protect personal data on the Web, actual online anonymity is quite limited. For instance, a user's Internet provider address, name and location can be monitored and exposed many times without the user's consent (Amichai-Hamburger and Perez, 2012).

Past research mainly explored users' online privacy in general (Nissenbaum, 2010; Solove, 2008) and found it to be quite low. In contrast, users expressed a high concern for their online privacy (Paine *et al.*, 2007; Wills and Zeljkovic, 2011). Women are generally more concerned with their privacy on the Web than men (Fogel and Nehmad, 2009; Graeff and Harmon, 2002; Hoy and Milne, 2010; Milne, Rohm and Bahl, 2004; O'Neill, 2001; Sheehan, 1999; Wills and Zeljkovic, 2011). Several studies have shown that although women tend not to disclose sensitive personal details, e.g., telephone numbers and home addresses (Acquisti and Gross, 2006; Feng and Xie, 2014; Fogel and Nehmad, 2009; Tufekci, 2008), they do disclose more personal details than men do (Hoy and Milne, 2010; Tufekci, 2008). Thus, despite having a high level of privacy concern, women rarely adopt the privacy protection behaviour that men do (Milne, Rohm and Bahl, 2004; Sheehan, 1999).

Some studies claim that this sex disparity originates in women's general lack of technological literacy (Park, 2015) and low digital self-efficacy (Hargittai and Shafer, 2006). Namely, even though women want to protect their online privacy, they do not possess sufficient knowledge or have confidence in their ability to do so. This explanation corresponds to one of Trepte *et al.*'s (2015) explanations of the privacy paradox-the knowledge gap hypothesis. According to this hypothesis, users lack of relevant literacy-in this case, online privacy literacy-makes it difficult for them to react in ways that reflect their attitudes and needs. Hence, increasing users' literacy levels of online privacy is important since it can help bridge the gap between privacy attitudes and related activities, as well as reduce privacy paradox behaviour (Trepte *et al.*, 2015).

The objective of this paper is to investigate the differences between men and women regarding the previously mentioned aspects of online anonymity and privacy. To model these differences and better understand the inter-influence of the different factors, we distinguish between technological dimensions and social dimensions in the examined factors' variables. Thus, we assess men and women's awareness of two types of threats against their online anonymity and privacy level: the technological threat (i.e., technology that enables surveillance, detection and exposure of a user's identity and personal details on the Web), and the social threat (i.e., exposure of a user's identity and personal details on the Web).

In addition, we examine the male/female differences regarding users' concern for the protection of their personal information on the Web and particularly on social networks. Furthermore, we explore the disparity in regard to levels of online privacy self-efficacy.

During the study, we provided a self-assessment to measure a user's ability to obtain the highest level of Web privacy and anonymity (Chen and Chen, 2015; Yao and Linz, 2008), and to measure online privacy literacy (Trepte *et al.*, 2015). To measure users' online privacy literacy, we estimated their familiarity level and actual usage of anonymity tools available on the Web. Based on Park's (2011, 2015) distinction, we considered online privacy literacy tools (e.g., cleaning the browser history), and privacy literacy skills used by social users (e.g., refraining from posting personal information).

Lastly, we investigated male/female differences in users' tendency to engage in privacy paradox behaviour (Barnes, 2006), i.e., users' preference to conveniently utilize the malleability of the Internet at the expense of information security, despite concern for their online privacy.

The following are specific research questions addressed in this study:

1. Are there differences in men's and womens' awareness of technological and social anonymity threats on the Web?
2. Are there differences in men's and womens' concern for the protection of personal information on social and non-social Websites?
3. Are there differences in men's and womens' online privacy self-efficacy and technological and social online privacy literacy?
4. Are there differences in men's and womens' tendency to engage in privacy paradox behaviour?
5. Does higher technological online privacy literacy decrease users' tendency to engage in privacy paradox behaviour?

To answer these questions, we conducted a user study with 169 Israeli academic students-71 men (42%) and 98 women (58%)-through a quantitative method using questionnaires with closed questions regarding user attitudes towards online privacy and anonymity. The questionnaires had over 40 questions.

This is the first attempt, to the best of our knowledge, to propose a comprehensive framework for empirical research of sex differences across various variables related to online privacy and anonymity. The study particularly integrates and examines variables that were not thoroughly explored in previous studies, such as awareness of two different types of anonymity threats, users' online privacy self-efficacy, and a tendency to engage in privacy paradox behaviour. In addition, this is the first study to investigate these issues among Israeli students from three different academic departments: information science, computer science, and accounting and business management. At the conceptual level, the significance of this study is the proposed framework and a comparative model of various factors based on the distinction between social and technological dimensions. This research has important social implications in the field of Internet education and well-being aimed at reducing the digital divide among the sexes, i.e., differences between sexes in regard to computer and Internet literacy, as well as self-efficacy.

The remainder of this paper is organized as follows. First, we review the related work. Second, we describe our research methodology. Third, we present the analysis and results. Finally, we summarise the study's main findings and contributions.

## Literature review

We first survey works related to the digital gap between the sexes, i.e., differences in computer and Internet literacy and self-efficacy. Further, we describe previous studies related to online privacy, anonymity and self-disclosure.

## Digital gap between the sexes

The digital divide is defined as the 'inequalities in access to the Internet, extent of use, knowledge of search strategies, quality of technical connections and social support, ability to evaluate the quality of information, and diversity of uses' (DiMaggio *et al.*, 2005). Wasserman and Richmond-Abbott (2005) distinguished between three aspects of this gap: access to the Internet, frequency of Internet use and scope of Internet use. Digital gap between the sexes is a widely explored topic; however, the findings of the different studies seem to be controversial.

On the one hand, many studies that investigated sex differences associated with technological aspects found a constant digital divide between men and women, which was noticeably reflected on the Internet (Bimber, 2000; Dixon *et al.*, 2014; Hargittai, 2010; Hargittai and Shafer, 2006; Hatlevik and Christophersen, 2013; Ono and Zavodny, 2003; Pan, Yan, Jing and Zheng, 2011; Tømte and Hatlevik, 2011; van Deursen and van Dijk, 2014, 2015; Wasserman and Richmond-Abbott, 2005; Weiser, 2000). A persisting gap between the sexes in terms of access and frequency of use was reported (Dixon *et al.*, 2014; Hargittai, 2010; Pan *et al.*, 2011; Weiser, 2000).

Studies conducted in Israel (Ganayem, Rafaeli and Azaiza, 2009; Lissitsa and Chachashvili-Bolotin, 2015) support these studies, while a national survey conducted by Israel's Ministry of Finance claims Israel's digital divide between the sexes is lower than Western Europe's (Ministry of Finance, 2005). Another study (Mesch and Talmud, 2011) found that Israeli men are more likely than women to express a positive attitude towards information and communications technology.

In addition, there are studies that report sexual inequality in online users' competence and skills (Hargittai, 2002, 2010; Tømte and Hatlevik, 2011; van Deursen and van Dijk, 2014, 2015).

On the other hand, some studies reveal that sexual inequalities in computer and Internet access are consistently narrowing (Bimber, 2000; Ono and Zavodny, 2003; Warf, 2013). Bimber (2000) measured sex differences in Internet access over a four-year period (1996 to 1999) and found that the fractions of men and women with access to the Internet have doubled. In 1996 and 1998, the sex gap for Internet access was only about 5%, which is not statistically significant. Ono and Zavodny (2003) state that there is no longer an online sex gap; however, they indicated that there continues to be a gap in frequency and intensity of use, although this gap is also diminishing. Likewise, Hatlevik and Christophersen (2013) did not find significant sex differences in digital competence.

Hargittai and Shafer (2006) also examined other aspects of the Internet sex gap-actual and perceived online skills. They found that there are no distinct sex gaps in online abilities. However, women's perceptions of their self-assessed skills, often termed as self-efficacy, are significantly lower than that of men. Another study (van Deursen and van Dijk, 2015) found that even though men appear to have better Internet skills than women, their actual online performances reveal that their digital competence is similar to that of women's in most cases.

## Sex differences in online information seeking behaviour

Online information seeking behaviour of men and women differs significantly in several aspects (Wasserman and Richmond-Abbott, 2005; Weiser, 2000). Weiser (2000) found that men use the Internet primarily for entertainment and leisure purposes, while women use it mostly for interpersonal communication and educational assistance. A later study (Choi *et al.*, 2009) revealed that men use the Internet in a more hedonic manner than women and often use it for recreational purposes. In addition, women used different types of Websites than men. Men were more likely to use financial, government, news and sexually explicit Websites, while women were more likely to use religious and culinary Websites (Wasserman and Richmond-Abbott, 2005). Men were more likely than women to purchase products and services online (Van Slyke, Comunale and Belanger, 2002; Shiu and Dawson, 2004; Zhang, Mandl and Wang, 2011; Lissitsa and Kol, 2016), even though this gap is narrowing (Faqih, 2016).

In addition, newer studies (e.g., Lee and Kim, 2014; Lee, Park and Hwang, 2015) report sex disparities in a new dimension of the user information behaviour in a mobile application setting. Lee and Kim (2014) found no significant sex differences in terms of mobile use, even though men were found to appreciate the aid of mobile media in their lives more than women. However, Lee, Park and Hwang (2015) did find sex disparity in terms of mobile utilization and particularly smartphone usage.

## Sex differences in online anonymity, privacy and self-disclosure

Park (2015) claims that the form of the disparity between the sexes in attitudes towards the Internet and in their online skills may have an influence on the issues of information protection and disclosure, and personal privacy management. User attitudes towards online privacy, anonymity and self-disclosure are widely explored topics in literature (Acquisti and Gross, 2006; Debatin, Lovejoy, Horn and Hughes, 2009; Dienlin and Trepte, 2015; Fogel and Nehmad, 2009; Graeff and Harmon, 2002; Hoy and Milne, 2010; Milne, Rohm and Bahl, 2004; Lee, Park and Kim, 2013; O'Neill, 2001; Paine *et al.*, 2007; Sheehan, 1999; Taddicken, 2014; Wills and Zeljkovic, 2011). Many of these studies conclude that most Internet users range from being concerned to very concerned regarding threats to their online privacy and anonymity, and are willing to take actions to protect it (e.g., Paine *et al.*, 2007; Wills and Zeljkovic, 2011). Nevertheless, most do not believe it is possible to be entirely anonymous online (Pew Research Center, 2014; Rainie, Kiesler, Kang and Madden, 2013).

Numerous studies conducted on online privacy and self-disclosure on the Web found sex as an influential factor. It was found that women are generally more concerned about their privacy on the Web than men (Graeff and Harmon, 2002; Milne, Rohm and Bahl, 2004; O'Neill, 2001; Sheehan, 1999; Taddicken, 2014; Wills and Zeljkovic, 2011), particularly on social networks (Fogel and Nehmad, 2009; Hoy and Milne, 2010). The main concern that discourages women from engaging in electronic commerce is the possible disclosure of personal information; however, men are willing to accept that risk for the earned profit (Michota, 2013).

Previously mentioned works explored users' concern with online privacy threat in general. However, in this paper we investigated the sex differences regarding two specific types of user anonymity threats: technology and social. We also assessed and compared men and women's concern for the protection of personal information on the Internet in general, and particularly on social networks.

Studies that were conducted within the social networks setting found that women tend to not disclose sensitive personal details, e.g., telephone numbers and home addresses (Acquisti and Gross, 2006; Feng and Xie, 2014; Fogel and Nehmad, 2009; Tufekci, 2008). However, women tend to disclose more non-sensitive personal details than men do (Hoy and Milne, 2010; Tufekci, 2008), e.g., their favourite books and movies, and information about their religion. In addition, they tend to post more photos than men (Kolek and Saunders, 2008). Two studies that were recently conducted in Israel in the setting of social networks showed that women tend to use their real name on their online profile and also post personal photographs (Bronstein, 2014); however, Zhitomirsky-Geffet and Bratspiess (2016) found no significant differences between the sexes in professional personal information disclosure on social networks.

Interestingly, despite women's high concerns about privacy, they rarely adopt privacy protection behaviour in contrast to men (Milne, Rohm and Bahl, 2004; Sheehan, 1999). A study found that even though women reported the same level of intention to adopt privacy protection strategies, they did not actually implement those intentions (Yao and Linz, 2008). This type of disparity between online privacy attitudes and behaviour has previously been termed the privacy paradox (Barnes, 2006; Norberg, Horne and Horne, 2007).

In this work, we examine the differences in a men's and women's tendency to privacy paradox behaviour, i.e. utilizing the malleability of the cyberspace at the expense of information security.

Park (2015, p. 2) asserted that '*Internet privacy makes gender disparity salient*'. As on the one hand, skills required for managing data may favour men who were found to be more technically proficient than women in various privacy tasks. On the other hand, privacy concern for data exposure may possibly sway women to exercise privacy skills that are more socially oriented. Park (2015) found that even though men were significantly more likely to embrace privacy protection behaviour involving the technical aspect, there were no sex differences in social behaviour related to privacy protection. Another study (Youn and Hall, 2008) showed that women do adopt social privacy protection strategies, but they do it differently than men. For example, female respondents protected themselves by providing inaccurate information as their privacy concerns increased, while male respondents refrained from registering on Websites.

In addition, our research explores sex differences regarding the technical dimensions of the online privacy literacy information system model. This is an attempt to extend Trepte *et al.*'s (2015) definition of two types of online privacy literacy: (1) passive online privacy literacy that comprises knowledge about technical aspects of online privacy and data protection, and (2) active online privacy literacy, which measures usage of tools and strategies for controlling online privacy. Furthermore, we distinguish between sex differences in technical online privacy literacy and social online privacy literacy level based on Park (2011, 2015). Similarly to previous research, the social online privacy literacy level is evaluated by providing inaccurate information and refraining from registering on Websites (Park, 2011; Youn and Hall, 2008). To assess the level of technical online privacy literacy, we used a list of the specific privacy control tools comprising a number of simple and more advanced techniques from existing literature (Park, 2011; Pew Research Center, 2014; Rainie *et al.*, 2013; Shelton, Rainie and Madden, 2015), which were not previously examined in the literature from the perspective of sex differences.

## Methods

### Sample Population

A sample of 169 students from three different Israeli academic departments was drawn: 1) accounting and business management studies in Bar-Ilan University, 2) information science studies in Bar-Ilan University and 3) computer science and engineering at the Jerusalem College of Technology.

As the sample population comprised only academia students, there is a relatively small age divergence. There was, however, equal sex distribution between different age groups, fields of study and education levels. Thus, demographic profiles and technical backgrounds of men and women could be considered similar and were not supposed to influence the inter-sex analysis presented in the next subsection. This study received ethics approval and was conducted in accordance with the American Psychology Association ethical requirements. It was made clear to the participants that the questionnaire was anonymous and would be used for research purposes only. Table 1 below presents the demographic characteristics of the sample.

| Variable | | Percentage | N |
|---|---|---|---|
| Sex | Male | 42 | 71 |
| | Female | 58 | 98 |
| Age | 21 | 31.36 | 53 |
| | 21-25 | 39.64 | 67 |
| | >25 | 28.99 | 49 |
| Education | Bachelor's degree | 88.8 | 150 |
| | Master's degree | 11.2 | 19 |
| Field of study | Accounting and business management | 23.67 | 40 |
| | Information science | 32.54 | 55 |
| | Computer science and engineering | 43.79 | 74 |

**Table 1: Demographic characteristics of the sample (n = 169)**

## Research variables and validation method

For purposes of this study, a questionnaire with closed questions was composed consisting of six groups of questions (items). The first group included demographic details of the participants (part A, items 1-3). The second group included indicators for measuring the level of technological and social anonymity threat awareness (part B, items 1-4). The third group included indicators for measuring the level of concern for the protection of personal information on general Websites (part B, items 5-6) and on social network sites (part B, items 7-8). The fourth group included an indicator for measuring users' online privacy self-efficacy levels (part B, item 9). The fifth group consisted of items composed to measure the level of technological and social online privacy literacy (part B, item 10; part C, items 1-16).

Technological online privacy literacy was measured as the knowledge and usage of privacy-enhancing tools (part C, items 1-16) based on Park's (2011) technical skills parameter, and on the list of tools examined in Rainie *et al.* (2013). An indicator for measuring the social online privacy literacy level was the tendency to avoid disclosing personal details or deliver falsified information while visiting a Website (part B, item 10), which is based on Park's (2011) social skills of privacy control.

The sixth group of items included indicators for measuring users' tendency to engage in privacy paradox behaviour, i.e., the preference of utilizing the malleability of cyberspace at the expense of information security (part D, items 1-25), which was based on Chellappa and Sin (2005), and information security surveys, such as Aydin and Chouseinoglou (2013) and Talib, Clarke and Furnell (2010).

Based on the described item groups above, the following research variables were defined:

1) Sex (part A, item 1) as an independent variable.

2) The level of anonymity threat awareness was measured using two different indicators for technological threat on anonymity: users' sense of anonymity while visiting a Website (part B, item 1) and users' awareness of the number of details that can be monitored while visiting a Website (part B, item 2). As opposed to the previously reviewed studies, which measured knowledge of institutional policies by generalized questions, we estimated users' awareness of many concrete parameters prone to online surveillance.

The participants were questioned regarding seven personal details that can be monitored while visiting a Website:

1. operating system
2. computer type
3. Web browser
4. Internet provider address
5. browsing history
6. location
7. name

Social threat awareness was assessed by users' sense of exposure to other users (part B, item 3).

Responses were coded as follows: 0 = no, 1 = yes. Lastly, a single value was calculated, which summed up the number of personal details that were marked by the participant.

3) The level of concern for the protection of personal information on general Websites and on social network sites was measured by two sets of indicators. Indicators for measuring concern on general Websites were the level of concern for the protection of personal information on the Web (part B, item 5), and the importance of protecting personal information on the Web (part B, item 6). Indicators for assessing concern on social networks were the level of concern for the protection of personal information on social networks (part B, item 7), and the importance of protecting personal information on social networks (part B, item 8). Two separate indices were then calculated by averaging each of these two sets of indicators respectively.

4) Users' online privacy self-efficacy levels were measured through one indicator that examined the belief in one's ability to browse anonymously (part B, item 9). The responses were coded on a 1-7 Likert scale and were then averaged.

5) The level of users' online privacy literacy was measured as follows:

a. Two different indicators measured the level of technological online privacy literacy: the knowledge level of privacy enhancing tools (part C, items 1-8), and the usage level of privacy enhancing tools (part C, items 9-16). The participants were questioned about eight different privacy-enhancing tools: (1) logging out from online accounts, (2) clearing the history and other browsing details, (3) blocking cookies, (4) browsing through incognito mode, (5) spoofing an internet protocol address, (6) using proxy servers, (7) using a virtual private network and (8) using third-generation onion routing.
   Each subject's responses were coded on a 1-5 Likert scale and then averaged over all the tools. A test of internal consistency and reliability (Cronbach's &alpha coefficient values) showed that the reliability of the indicator measuring the level of knowledge of privacy-enhancing tools in the present sample was ? = 0.86. The test also showed that the reliability for the indicator measuring the level of usage of privacy-enhancing tools was ? = 0.78.
   A Pearson correlation test revealed a positive and strong correlation

(r = 0.76, p = 0.001) between the level of passive online privacy literacy (knowledge of privacy-enhancing tools) and the level of active online privacy literacy (using privacy-enhancing tools).

b. The social online privacy literacy was measured through one binary indicator (part B, item 10), reflecting the tendency to refrain from submitting personal details or submitting falsified information while visiting a Website. All responses were coded with the values of 0 = no and 1 = yes.

We conclude that the internal consistencies of the above measures assessed by means of the Cronbach's alpha coefficient were above 0.70, and thus can be considered acceptable (Nunnally and Bernstein, 1994).

## Results

In this section, we present the results of the statistical analysis to test the research questions presented above. First, we analysed sex differences with regard to general awareness of anonymity limitations on the Web. Table 2 presents the range, average and standard deviation of sex differences with regard to the four anonymity threat awareness indicators from the sample. Table 2 reports that there were significant differences between men and women for all of the examined anonymity threat awareness indicators. On the one hand, women reported a lower awareness level of technological threats than men did. As a result, women sensed a higher anonymity level while visiting a Website (2.66 versus 2.27), and lower awareness levels of monitored personal details compared to men (3.23 versus 4.52). On the other hand, women reported a higher level of social threat awareness than men did, as they felt a significantly higher level of exposure to other users than men did (3.18 versus 2.94).

| Anonymity threat awareness indicators | Men M (SD) | Women M (SD) | T df = 167 | P |
|---|---|---|---|---|
| The sense of anonymity while visiting a Website | 2.27 (0.88) | 2.66 (0.94) | 2.73 | 0.01 |
| Awareness of the number of details that can be monitored on a Website | 4.52 (2.01) | 3.23 (2.09) | 4.02 | 0.001 |
| The sense of exposure to other users | 2.94 (0.83) | 3.18 (0.82) | 1.88 | 0.03 |

**Table 2: The range, average and standard deviation of the sex differences with regard to technological and social anonymity threat awareness indicators of the sample (n = 169)**

There were no significant differences between men and women in regard to the level of concern for protecting personal information on the Web in general.

Subsequently, we analysed sex differences in regard to the level of concern for protecting personal information on social network sites. Table 3 presents the range, average and standard deviation of the sex differences with regard to the level of concern for protecting personal information on social networks of the sample.

| | | | T | |
|---|---|---|---|---|

| | Men M (SD) | Women M (SD) | df = 163 | P |
|---|---|---|---|---|
| Privacy concern on social networks | 3.51 (0.93) | 3.82 (1.07) | -1.98 | 0.049 |

**Table 3: The range, average and standard deviation of the sex differences with regard to the level of concern for protecting personal information on social networks of the sample (n = 169)**

Women were found to be more concerned than men with protecting personal information on social networks (3.82 versus 3.51), as shown in Table 3. No significant sex differences were found in the level of online privacy self-efficacy (3.66 versus 3.48 on average, respectively).

Table 4 presents the range, average and standard deviation of sex differences for the technological online privacy literacy level of the sample.

| | Men M (SD) | Women M (SD) | T df = 165 | P |
|---|---|---|---|---|
| Passive online privacy literacy | 3.15 (0.97) | 2.55 (0.85) | 4.24 | 0.001 |
| Active online privacy literacy | 2.50 (0.85) | 2.06 (0.69) | 3.74 | 0.001 |
| Average online privacy literacy | 2.83 | 2.31 | | |

**Table 4: The range, average and standard deviation of the sex differences with regard to the passive and active technological online privacy literacy level of the sample (n = 169)**

Table 4 shows that compared to men, there were significantly lower knowledge levels (passive online privacy literacy) and use (active online privacy literacy) of privacy-enhancing tools for women (2.55 versus 3.15, and 2.50 versus 2.06, respectively).

With regard to the social online privacy literacy, we found that slightly less than half of the sample (48%) reported that they do avoid disclosing personal details or delivering falsified information while visiting a Website. However, no significant difference was detected for this variable's distribution between the sexes, which was virtually equal for men and women. Lastly, we analysed sex differences in regard to the tendency to engage in privacy paradox behaviour. No significant differences were found between men and women for this variable (2.77 versus 2.68 on average, respectively).

## Discussion and conclusions

In this paper, we investigated male/female disparity with regard to various aspects of online privacy and anonymity. The proposed comparative model summarizing users' attitudes toward online privacy and anonymity is displayed in Figure 1. Factors with significantly higher values for men are shown on the left side of the diagram. Factors with significantly higher values for women are shown on the right side of the diagram, while factors with similar values for both sexes are shown in the middle. The arrows show the expected influences between the different factors based on current and previous studies (Chen and Chen, 2015; Milne, Rohm and Bahl, 2004; Park, 2011, 2015; Sheehan, 1999; Yao and Linz, 2008).

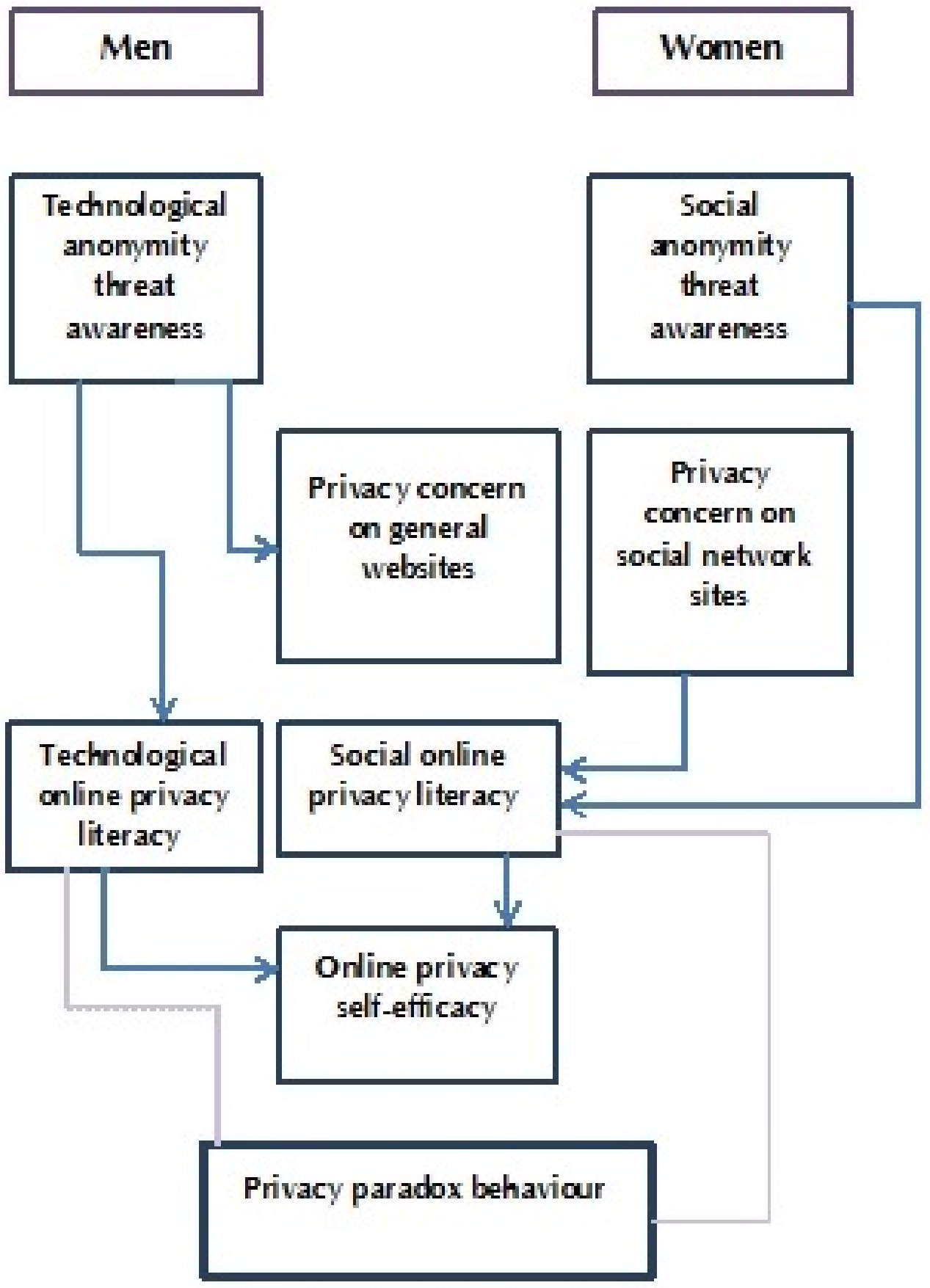

**Figure 1: The integrative comparative model of sex disparities in users' online privacy and anonymity attitudes.**

At first, we analysed sex differences with regard to the general awareness of limited anonymity on the Web by using several measures of technological and social threat awareness. Compared to men, women were found to sense a higher anonymity level while visiting a Website, and a lower awareness level of monitored personal details. However, they felt a higher level of exposure to other users. Namely, women showed low technological threat awareness and high social threat awareness when compared to men. These findings might be related to women's lower levels of trust regarding social networking websites (Fogel and Nehmad, 2009).

We then analysed sex differences with regard to the level of concern for protecting personal information on general and social network Websites. As opposed to previous works (Graeff and Harmon, 2002; Milne, Rohm and Bahl, 2004; O'Neill, 2001; Sheehan, 1999; Wills and Zeljkovic, 2011), no significant differences were found regarding the level of concern for protecting personal information on general Websites. However, women were found to be more concerned than men about protecting their personal information on social networks. The latter finding corresponds with previous studies (Fogel and Nehmad, 2009; Hoy and Milne, 2010) that found women had a higher concern for their privacy on social network sites than men did.

Further, we analysed sex differences in users' levels of online privacy self-efficacy and online privacy literacy (technical and social). As opposed to a previous study (Hargittai and Shafer, 2006) that found women's online self-efficacy to be significantly lower than men's online self-efficacy, we found no significant sex differences for this variable. Nevertheless, the study did reveal significant differences between men and women in the levels of technical online privacy literacy, as women showed lower levels of both active and passive online privacy literacy. These differences can be explained by a lower level of Internet literacy among women, and might reflect a digital gap that exists between the sexes, as reported in previous research (Bimber, 2000; Ono and Zavodny, 2003; Warf, 2013). A high correlation between knowledge and usage of privacy protection tools shows the importance of knowledge regarding protection of anonymity and privacy on the Web. The higher the user's knowledge of the tools is, the higher his or her intent and ability to use them to protect online personal information is. Yet, in contrast to some previous studies (e.g., Milne, Rohm and Bahl, 2004; Sheehan, 1999; Yao and Linz, 2008), no significant differences in social online privacy literacy between men and women were found.

Lastly, we analysed sex differences with regard to privacy paradox behaviour. In contrast to previous studies (Hoy and Milne, 2010; Milne, Rohm and Bahl, 2004; Sheehan, 1999; Tufekci, 2008; Yao and Linz, 2008), no significant differences were found between men and women for this variable. Thus, our findings do not support Trepte *et al.*'s (2015) hypothesis, and shows that a higher technical online privacy literacy level possessed by men does not necessarily decrease their tendency toward personal information disclosure and privacy paradox behaviour as compared to women. We concluded that some other factors might influence users' tendency toward privacy paradox behaviour. Therefore, we noted the relationship between users' online privacy literacy and privacy paradox behaviour with a light purple hyphen-line on the diagram in Figure 1. Further exploration of influential factors of privacy paradox behaviour is subject for future work.

In summary, the findings of this research refined those of previous studies regarding sex differences in users' attitudes to online privacy and anonymity. Previous studies, as mentioned above, found that women are more concerned with their privacy (on the Web in general and on social networks) than men, disclose more personal information, engage more in privacy paradox behaviour, and have a lower online privacy literacy and online privacy self-efficacy level. Our results, which were based on Israeli students' population analysis, revealed that women reported a higher awareness and concern only for social threats on their privacy, while being similar to men concerning social online privacy literacy, online privacy self-efficacy, a tendency to personal information disclosure and privacy paradox behaviour. In accordance with the findings of Park (2015), we found that the only two factors for which men still have an advantage over women were related to technology: awareness of the technological threat and the technological online privacy literacy level. Thus overall, the divide between the sexes in terms of attitudes towards online anonymity and privacy protection has been reduced in several respects.

This study's findings are limited since the study population was only composed of students from three specified academic departments, and are based on respondents' self-reported skills and behaviour. Therefore, a further qualitative research of sex disparities concerning various factors of users' online privacy behaviour is required. In addition, future work should apply qualitative analysis to explore additional types and affecting factors of online privacy behaviour. Most students today are digitally oriented and proficient in utilizing online tools, which might have an effect on the findings. Therefore, further research aiming to generalize the paper results should apply the proposed methodology on additional population types comprising subjects with different age groups, education levels, occupation types, cultural backgrounds and countries of origin.

## Social significance and implications

Our results have high social significance. It can be expected that a higher online privacy self-efficacy, anonymity threat awareness and privacy concern, and a higher level of online privacy literacy will reduce personal information disclosure and lead to higher identity and personal data protection of Web users. However, we found that women's relatively high online privacy self-efficacy level, which is probably based on their low level of technological threat awareness, does not match their relatively low technological online privacy literacy level. This leads to a lower ability to protect their identity and personal information as compared to men. Conversely, men's technological threat awareness, which is higher than their online privacy self-efficacy, along with their relatively higher online privacy literacy provides them with an increased ability to protect their identity and personal data. Women's ability to effectively manage their online privacy in the digital age is crucial for various tasks (Park, 2015), such as health, educational, financial, commercial, social and political information seeking and consumption. We also suggest that our methodological distinction between social and technological threats and social and technological online privacy literacy be applied to the wider contexts of cyber security investigation as a basis for developing new measures to protect personal data. Thus, users' awareness of technological threats should be increased as part of the improvement of online information literacy. Furthermore, in view of modern phenomena that are changing online behaviour in very radical ways, there is a growing concern for online privacy protection among privacy commentators and law enforcement agencies. Therefore, the social implication of this research is that further steps, including policy intervention and educational programmes, need to be taken to eliminate the inter-sex technological gap in online privacy and anonymity awareness and literacy.

## About the author

**Maor Weinberger** is a Ph.D. student in the Department of Information Science, Bar-Ilan University, Israel. He can be contacted at:

maor89@gmail.com

**Maayan Zhitomirsky-Geffet** is a lecturer and assistant professor in the Department of Information Science, Bar-Ilan University, Israel. She can be contacted at: Maayan.Zhitomirsky-Geffet@biu.ac.il

**Dan Bouhnik** is a lecturer in the Department of Information Science, Bar-Ilan University, Israel and in the Department of Computer Science, Jerusalem College of Technology, Israel. He can be contacted at: bouhnik@jct.ac.il

## References

Acquisti, A. & Gross, R. (2006). Imagined communities: awareness, information sharing, and privacy on the Facebook. In *The 6th International Workshop on Privacy Enhancing Technologies.* Cambridge, UK.

Amichai-Hamburger, Y. & Perez, A. (2012). Anonymiut veinteraktiviut bainterenet: hazchut lepratiut kemusag rav memadi [Anonymity and interactivity on the Internet: the right to privacy as a multi-dimensional concept]. In T. Schwartz-Altshuler (Ed.), *beidan shel shinui [Privacy in the age of change]*, (pp. 201-230). Jerusalem, Israel: Israel Democracy Institute.

Aydin, O. M. & Chouseinoglou, O. (2013). Fuzzy assessment of health information system users' security awareness. *Journal of Medical Systems, 37*(6).

Barnes, B. S. (2006). A privacy-paradox: social networking in the United States. *First Monday, 11*(9). Retrieved from http://firstmonday.org/article/view/1394/1312.

Bimber, B. (2000). Measuring the gender gap on the Internet. *Social Science Quarterly, 81*(3), 868-876. Retrieved from http://www.dleg.state.mi.us/mpsc/electric/workgroups/lowincome/internet_gender_gap.pdf.

Bronstein, J. (2014). Creating possible selves: information disclosure behaviour on social networks. *Information Research, 19*(1) paper 609. Retrieved from http://www.informationr.net/ir/19-1/paper609.html.

Chellappa, R. K. & Sin, R. G. (2005). Personalization versus privacy: an empirical examination of the online consumer's dilemma. *Information Technology & Management, 6*(2-3), 181-202.

Chen, H. T. & Chen, W. H. (2015). Couldn't or wouldn't? The influence of privacy concerns and self-efficacy in privacy management on privacy protection. *Cyberpsychology, Behaviour and Social Networking, 18*(1), 13-19.

Choi, K., Son, H., Park, M., Han, J., Kim, K., Lee, B. & Gwak, H. (2009). Internet overuse and excessive daytime sleepiness in adolescents. *Psychiatry and Clinical Neurosciences, 63*(4), 455-462.

DiMaggio, P., Hargittai, E., Neuman, W. R. & Robinson, J. P. (2001). Social implications of the Internet. *Annual Review of Sociology, 27,* 307-336.

Debatin, B., Lovejoy, J. P., Horn, A. K. & Hughes, B. N. (2009). Facebook and online privacy: attitudes, behaviours and unintended consequences. *Journal of Computer-Mediated Communication, 15*(1), 83-108.

Dienlin, T. & Trepte, S. (2015). Is the privacy paradox a relic of the past?

An in-depth analysis of privacy attitudes and privacy behaviours. *European Journal of Social Psychology, 45*(3), 285-297.

Dixon, L. J., Correa, T., Straubhaar, J., Covarrubias, L., Graber, D., Spence, J. & Rojas, V. (2014). Gendered space: the digital divide between male and female users in Internet public access sites. *Journal of Computer-Mediated Communication, 19*(4), 991-1009.

Faqih, K. M. S. (2016). An empirical analysis of factors predicting the behavioural intention to adopt Internet shopping technology among non-shoppers in a developing country context: does gender matter? *Journal of Retailing and Consumer Services, 30*, 140-164.

Feng, Y. & Xie, W. (2014). Teens' concern for privacy when using social networking sites: an analysis of socialization agents and relationships with privacy protecting behaviours. *Computers in Human Behaviour, 33*, 153-162.

Fogel, J. & Nehmad, E. (2009). Internet social network communities: risk taking, trust and privacy concerns. *Computers in Human Behaviour, 25*(1), 153-160.

Ganayem, A., Rafaeli, S. & Azaiza, F. (2009). Pa'ar digitali: hashimush beinternet bachevra ha'arvit beisrael [Digital divide: Internet usage within Israeli Arab society]. *Megamot, 46(1-2)*, 164-196.

Graeff, T. R. & Harmon, S. (2002). Collecting and using personal data: consumers' awareness and concerns. *Journal of Consumer Marketing, 19(4-5)*, 302-318.

Hargittai, E. (2002). Second-level digital divide: differences in people's online skills. *First Monday, 7*(4). Retrieved from http://firstmonday.org/article/view/942/864.

Hargittai, E. (2010). Digital na(t)ives? Variation in Internet skills and uses among members of the "net generation". *Sociological Inquiry, 80*(1), 92-113.

Hargittai, E. & Shafer, S. (2006). Differences in actual and perceived online skills: the role of gender. *Social Science Quarterly, 87*(2), 432-448.

Hatlevik, O. E. & Christophersen, K-A. (2013). Digital competence at the beginning of upper secondary: identifying factors explaining digital inclusion. *Computers & Education, 63*, 240-247.

Hoy, M. G. & Milne, G. (2010). Gender differences in privacy-related measures for young adult Facebook users. *Journal of Interactive Advertising, 10*(2), 28-45.

Kolek, E. A. & Saunders, D. (2008). Online disclosure: an empirical examination of undergraduate Facebook profiles. *Journal of Student Affairs Research and Practice, 45*(1), 1-25.

Lee, J. H. & Kim, J. (2014). Socio-demographic gaps in mobile use, causes and consequences: a multi-group analysis of the mobile divide model. *Information, Communication & Society, 17*(8), 917-936.

Lee, H., Park, N. & Hwang, Y. (2015). A new dimension of the digital divide: exploring the relationship between broadband connection, smartphone use and communication competence. *Telematics and Informatics, 32*(1), 45-56.

Lee, H., Park, H. & Kim, J. (2013). Why do people share their context information on social network services? A qualitative study and an experimental study on users' behaviour of balancing perceived benefit and risk. *International Journal of Human-Computer Studies, 71*(9), 862-877.

Lissitsa, S. & Chachashvili-Bolotin, S. (2015). Does the wind of change blow in late adulthood? Adoption of ICT by senior citizens during the past decade. *Poetics, 52*, 44-63.

Lissitsa, S. & Kol, O. (2016). Generation X vs. Generation Y-a decade of online shopping. *Journal of Retailing and Consumer Services, 31*, 304-312.

Mesch, G. S. & Talmud, I. (2011). Ethnic differences in Internet access: the role of occupation and exposure. *Information, Communication & Society*, *14*(4), 445-471.

Michota, A. (2013). Digital security concerns and threats facing women entrepreneurs. *Journal of Innovation and Entrepreneurship, 2*(7), 1-11.

Milne, G. R., Rohm, A. J. & Bahl, S. (2004). Consumers' protection of online privacy and identity. *Journal of Consumer Affairs, 38*(2), 217-232.

Ministry of Finance. (2005). *Seker muchanut upe'arim digitalim 2005 [E-readiness and digital divide survey 2005].* Jerusalem, Israel: Ministry of Finance.

Nissenbaum, H. (2010). *Privacy in context: technology, policy, and integrity of social life.* Stanford, CA: Stanford University Press.

Norberg, P. A., Horne, D. R. & Horne, D. A. (2007). The privacy paradox: personal information disclosure intentions versus behaviours. *Journal of Consumer Affairs, 41*(1), 100-126.

Nunnally, J. & Bernstein, I. (1994). *Psychometric theory (3rd ed.)*. New York, NY: McGraw-Hill.

O'Neill, D. (2001). Analysis of Internet users' level of online privacy concerns. *Social Science Computer Review, 19*(1), 17-31.

Ono, H. & Zavodny, M. (2003). Gender and the Internet. *Social Science Quarterly, 84*(1), 111-121.

Paine, C., Reips, U. D., Steiger, S., Joinson, A. & Buchanan, T. (2007). Internet users' perceptions of "privacy concerns" and ""privacy actions". *International Journal of Human-Computer Studies, 65*(6), 526-536.

Pan, Z., Yan, W., Jing, G. & Zheng, J. (2011). Exploring structured inequality in Internet use behaviour. *Asian Journal of Communication, 21*(2), 116-132.

Park, Y. J. (2011). Digital literacy and privacy behaviour online. *Communication Research*, *40*(2), 215-236.

Park, Y. J. (2015). Do men and women differ in privacy? Gendered privacy and (in)equality in the Internet. *Computers in Human Behavior, 50*, 252-258.

Pew Research Center. (2014). Public perceptions of privacy and security in the post-Snowden era. Retrieved from http://www.pewInternet.org/files/2014/11/PI_PublicPerceptionsofPrivacy_111214.pdf.

Pfitzmann, A. & Köhntopp, M. (2001). Anonymity, unobservability and pseudonymity-a proposal for terminology. In H. Federrath, (Ed.), *Proceedings of the International Workshop on Design Issues in Anonymity and Unobservability* (pp. 1-9). Berlin: Springer-Verlag.

Rainie, L., Kiesler, S., Kang, R. & Madden, M. (2013). Anonymity, privacy and security online. Retrieved from http://www.pewinternet.org/~/media//Files/Reports/2013/PInternet provider_AnonymityOnline_090513.pdf.

Sheehan, K. B. (1999). An investigation of gender differences in online privacy concerns and resultant behaviors. *Journal of Interactive*

*Marketing, 18*(4), 24-38.

Shelton, M., Rainie, L. & Madden, M. (2015). American's privacy strategies post-Snowden. Retrieved from http://www.pewInternet.org/files/2015/03/PI_AmericansPrivacyStrategies_0316151.pdf.

Shiu, E. C. C. & Dawson, J. A. (2004). Comparing the impacts of Internet technology and national culture on online usage and purchase from a four-country perspective. *Journal of Retailing and Consumer Services, 11*(6), 385-394.

Solove, D. J. (2008). *Understanding privacy*. Cambridge, MA: Harvard University Press.

Taddicken, M. (2014). The "privacy paradox" in the social web: the impact of privacy concerns, individual characteristics, and the perceived social relevance on different forms of self-disclosure. *Journal of Computer-Mediated Communication, 19*(2), 248-273.

Talib, S., Clarke, N. L. & Furnell, S. M. (2010). An analysis of information security awareness within home and work environments. In *Proceedings of the 5th International Conference on Availability, Reliability and Security*. Krakow, Poland: Conference Publishing Services.

Tømte, C. & Hatlevik, O. E. (2011). Gender-differences in self-efficacy ICT related to various ICT-user profiles in Finland and Norway. How do self-efficacy, gender and ICT-user profiles relate to findings from PISA 2006. *Computers & Education, 57*(1), 1416-1424.

Trepte, S., Teutsch, D., Masur, P. K., Eicher, C., Fischer, M., Hennhöfer, A. & Lind, F. (2015). Do people know about privacy and data protection strategies? Towards the online privacy literacy scale (OPLIS). In S. Gutwirth, R. Leenes & P. de Hert (Eds.), *Reforming European data protection law* (pp. 333-365). Dordrecht, Netherlands: Springer.

Tufekci, Z. (2008). Can you see me now? Audience and disclosure regulation in online social network sites. *Bulletin of Science Technology Society, 28*(1), 20-36.

van Deursen, A. J. A. M. & van Dijk, J. A. G. M. (2014). The digital divide shifts to differences in usage. *New Media & Society, 16*(3), 507-526.

van Deursen, A. J. A. M. & van Dijk, J. A. G. M. (2015). Internet skill levels increase, but gaps widen: a longitudinal cross-sectional analysis (2010-2013) among the Dutch population. *Information, Communication and Society, 18*(7), 782-797.

Van Slyke, C., Comunale, C. L. & Belanger, F. (2002). Gender differences in perceptions of Web-based shopping. *Communications of the Association for Computing Machinery, 45*(8), 82-86.

Warf, B. (2013). Contemporary digital divides in the United States. *Journal of Economic and Social Geography, 104*(1), 1-17.

Wasserman, I. M. & Richmond-Abbot, M. (2005). Gender and the Internet: causes of variation in access, level and scope of use. *Social Science Quarterly, 86*(1), 252-270.

Weiser, E. B. (2000). Gender differences in Internet use patterns and Internet application preferences: a two-sample comparison. *CyberPsychology & Behaviour, 3*(2), 167-178.

Wills, C. E. & Zeljkovic, M. (2011). A personalized approach to Web privacy: awareness, attitudes and actions. *Information Management & Computer Security, 19*(1), 53-73.

Yao, M. Z. & Linz, D. G. (2008). Predicting self-protections of online privacy. *Cyberpsychology & Behavior, 11*(5), 615-617.

Youn, S. & Hall, K. (2008). Gender and online privacy among teens: risk perception, privacy concerns and protection behaviours. *Cyberpsychology & Behavior*, *11*(6), 763-765.

Zhang, J., Mandl, H. & Wang, E. (2011). The effect of vertical-horizontal individualism-collectivism on acculturation and the moderating role of gender. *International Journal of Intercultural Relations*, *35*(1), 124-134.

Zhitomirsky-Geffet, M. & Bratspiess, Y. (2016). Professional information disclosure on social networks: the case of Facebook and LinkedIn in Israel. *Journal of the Association for Information Science and Technology*, *67*(3), 493-504.

---

## How to cite this paper

# Appendix A

## Questionnaire

### Part A - Demographic Details

1. Sexß: M or F
2. Year of Birth: ______
3. Education:
    a. Bachelor's degree
    b. Master's degree
    c. Doctor of philosophy
    d. Other, please specify ____________

    Academic institution: _________
    Field of study: __________

### Part B

1. How anonymous do you feel while surfing the Web? (Please circle)
    a. not anonymous at all
    b. partially anonymous
    c. moderately anonymous
    d. highly anonymous
    e. very highly anonymous
2. When you visit a Website, which of the following can the Website determine? (Multiple choices are permitted)
    a. your operating system
    b. your computer type

c. your Web browser
    d. your Internet provider address
    e. your browsing history
    f. your location
3. In your opinion, how exposed are you to other users on the Internet? (Please circle)
    a. completely exposed
    b. highly exposed
    c. moderately exposed
    d. lowly exposed
    e. not exposed at all
4. When you visit a Website, which of the following can another user determine? (Multiple choices are permitted)
    a. your operating system
    b. your computer type
    c. your Web browser
    d. your Internet provider address
    e. your browsing history
    f. your location
5. How concerned are you about the protection of your personal information on the Web? (Please circle)
    a. not concerned at all
    b. slightly concerned
    c. moderately concerned
    d. highly concerned
    e. very highly concerned
6. How important is it for you to protect your personal information when surfing a Website? (Please circle)
    a. not important at all
    b. slightly important
    c. moderately important
    d. highly important
    e. very highly important
7. How concerned are you about the protection of your personal information when using social network sites? (Please circle)
    a. not concerned at all
    b. slightly concerned
    c. moderately concerned
    d. highly concerned
    e. very highly concerned
8. How important it is for you to protect your personal information on social networks? (Please circle)
    a. not important at all
    b. slightly important
    c. moderately important
    d. highly important
    e. very highly important
9. What is your level of belief in your own ability to browse the Web

anonymously, if necessary? (Please circle)

a. no belief at all
b. low level of belief
c. low to moderate level of belief
d. moderate level of belief
e. moderate to high level of belief
f. high level of belief
g. very high level of belief

10. In many Websites, the user is required to submit personal details (e.g., name, telephone number, email address, etc.) to obtain various services that the Website provides. On most occasions, do you tend to submit falsified information or refrain from submitting personal information when asked to do so?

    Yes or No

## Part C

1-8. What is your knowledge level of each of the following privacy-enhancing tools? (Please circle for each tool)

Possible answers for every separate tool were on the 1-5 Likert scale: 1) no knowledge of, 2) low knowledge level, 3) moderate knowledge level, 4) high knowledge level and 5) very high knowledge level.

a. logging out from online accounts
b. clearing history and other browsing details
c. blocking cookies
d. browsing through incognito mode
e. spoofing the Internet provider
f. using proxy servers
g. using virtual private networks
h. using the onion routing

9-16. What is your level of usage of each of the following privacy-enhancing tools? (Please circle for each tool)

The possible answers for every tool were the following: 1) no usage at all, 2) low usage level, 3) moderate usage level, 4) high usage level and 5) very high usage level.

a. logging out from online accounts
b. clearing history and other browsing details
c. blocking cookies
d. browsing through incognito mode
e. spoofing the Internet provider
f. using proxy servers
g. using virtual private networks
h. using the onion routing

## Part D

Please circle your level of agreement with each of the following statements.

The possible answers for every statement were the following: 1) do not agree at all, 2) lowly agree, 3) moderately agree, 4) highly agree and 5) very highly agree.

1. I use complicated passwords, even though it takes me more time, to reduce the risk for my personal information to be stolen on the Internet.
2. I do not tend to use the save password option when it is offered to me by the Web browser.
3. I often change the passwords for my online accounts, even though it may be tedious.
4. I am willing to pay for services that will guarantee the protection of my personal information.
5. I think that online privacy-enhancing services flaw my surfing experience.
6. I tend to read the privacy policy statement of a Website asking me to submit personal details.
7. I tend to download software and content that I find to be important, even from unfamiliar Websites.
8. I do not tend to conduct online shopping, because I am concerned for my information security.
9. I am willing to submit personal information to Websites to get online advertisements that are customized to my personal interests.
10. I am willing to submit personal information to social networks applications to get messages and services that are customized to my personal interests.
11. I tend to use a simple password that is easy to remember when I am required to set a new one.
12. I use online banking services, and not only for checking my account's status.
13. I do not tend to use the same password for different online accounts.
14. I tend to install different extensions on my Web browser, even when it requires me to submit personal details.
15. I seldom change the access passwords for my online accounts.
16. I am willing to submit personal information on Websites for the purpose of online advertising in exchange for monetary compensation.
17. I am willing to disclose personal information on social networks to gain better social interaction, social endorsement, and also to receive interesting services and information, or any other benefit.
18. In general, I prefer to comfortably use the Internet, even at the expense of protecting my personal information.
19. I do not tend to read the privacy policy statement of a Website I am visiting.
20. I often tend to conduct online shopping, even though I know it might jeopardize my information security.
21. I tend to download software and services aimed to improve the performance of my computer and/or Web browser, even from

seemingly unprotected Websites.

22. I will not submit personal information on an unsecured Website, even if it offers me a service I desire.
23. I tend to open pop-ups only when they are dealing with a matter of special personal interest.
24. I tend to visit Websites that interest me, even though I know for certain they are using cookies for the purpose of personalized advertising.
25. In general, I prefer to protect my information security, even at the expense of my comfortable use of the Internet.

---

---

---